\documentclass[%
reprint,
superscriptaddress,
showpacs,preprintnumbers,
showkeys,
 amsmath,amssymb,
 aps,
]{revtex4-1}

\usepackage{graphicx}
\usepackage{dcolumn}
\usepackage{bm}

\def\be{\begin{equation}}
\def\ee{\end{equation}}
\def\bea{\begin{eqnarray}}
\def\eea{\end{eqnarray}}
\newcommand{\ket}[1]{\mbox{$|#1\rangle$}}
\newcommand{\bra}[1]{\mbox{$\langle#1|$}}
\newcommand{\avg}[1]{\mbox{$\langle#1\rangle$}}

\begin{document}

\title {Strong Coupling between a Topological Qubit and a Nanomechanical Resonator}

\author{Fang-Yu Hong}
\affiliation{Department of Physics, Center for Optoelectronics Materials and Devices, Zhejiang Sci-Tech University,  Hangzhou, Zhejiang 310018, China}
\author{Jing-Li Fu}
\affiliation{Department of Physics, Center for Optoelectronics Materials and Devices, Zhejiang Sci-Tech University,  Hangzhou, Zhejiang 310018, China}
\author{Zhi-Yan Zhu}
\affiliation{Department of Physics, Center for Optoelectronics Materials and Devices, Zhejiang Sci-Tech University,  Hangzhou, Zhejiang 310018, China}
\date{\today}
\begin{abstract}
 We describe a scheme that enables a strong coherent coupling between a topological qubit and the quantized motion of a magnetized nanomechanical resonator. This coupling is achieved by attaching an array of magnetic tips to a namomechanical resonator under a quantum phase controller which coherently controls  the energy gap of a topological qubit.
     Combined with single-qubit rotations the strong coupling enables arbitrary unitary transformations  on  the  hybrid system of  topological and mechanical qubits  and may pave the way for the quantum information transfer between topological and optical qubits. Numerical simulations show that quantum state transfer and  entanglement distributing between the topological and mechanical qubits may be accomplished with high fidelity.
\end{abstract}


\maketitle
{\it Introduction.}---A major challenge facing the field of quantum information processing (QIP) arises from the delicate nature of a quantum system, their tendency to decohere into classical states through coupling to the environment.
 To address this obstacle there emerged  some interesting topological quantum computation schemes \cite{ayki,cnsh}, where quantum information is stored in nonlocal (topological) degrees of freedom of topologically ordered systems. Being decoupled from local perturbations these nonlocal degrees of freedom  enable the topological QIP approaches  to obtain its extraordinary fault tolerance and to have a huge advantage over conventional ones.  As the simplest non-Abelian excitation for topological qubits, the zero energy Majorana bound state (MBS) \cite{fwil}, is conjectured to be  exist in the spin lattice systems \cite{ayki}, in the $p+ip$ superconductors \cite{ nrdg}, in the filling fraction $\nu=5/2$ fractional quantum Hall system \cite{cnsh}, in the superconductor Sr$_2$RuO$_4$ \cite{sscn}, in the   topological insulators coupled to s-wave superconductors \cite{lfck,mhck}, and in some semiconductors of strong spin-orbit interaction coupled to superconductors \cite{jsrl,jali,yogr,rljs} where an experimental observation has recently verified its existence \cite{vmou}.

On the other hand, the nonlocal nature of  topological qubits makes it tough to measure and manipulate  them, because they can only be controlled by globe braiding operations, i.e., by physical exchange of the associated local non-Abelian anyons \cite{aste,daiv}. Furthermore these braiding operations  for Ising anyons alone are  not sufficient to accomplish  universal quantum computation and have to be combined with topologically unprotected operations \cite{pbon,pbond}.  Implementing unprotected operations within a topological system proves to be very challenging due to the existence of  significant nonuniversal effects \cite{pbrl}. At the same time, stead advancements have been achieved in conventional QIP systems, such as the recent progresses in a basic quantum network of single atoms in optical cavities \cite{srcn}, in long lifetime of nuclear spins in a diamond crystal \cite{pmgk,mglc}, in  high fidelity operations on trapped ions \cite{rbdw} and on superconducting qubits \cite{jcfw}, in distributing  entanglement between single-atoms at a distance \cite{dmpm} and between an optical photon and a solid-state spin qubit \cite{etyc}.

  Thus the best solution is to make hybrid systems by combining the advantages of  topological qubits, robust quantum storage and protected gates,  with those of conventional qubits such as high fidelity readout, universal gates, and  quantum network. Such hybrid schemes have recently been suggested for the  anyons coupled to superconducting flux qubits \cite{fhaa,jsst,ljck} and for the anyons in atomic spin lattices \cite{ljia}, in  optical lattices \cite{magu}, and in Majorana nanowires \cite{aykit} coupled to a semiconductor double-dot qubit \cite{pbrl}.
Here we propose a scheme for quantum information transfer between a magnetized nanomechanical resonator \cite{mphj,prpc, ahgs} and a topological qubit  encoded on  Majorana fermions (MFs) on the surface of a topological insulator (TI) \cite{lfck}. The motion of the resonator under a quantum phase-controller (QPC) \cite{ljclk} modifies the energy gap between the two topological qubit states, resulting in  a strong coupling between  the topological qubit and the quantized motion of the resonator with its strength conveniently controlled by the QPC.
    Based on this strong coupling arbitrary quantum information transfer and quantum entanglement distribution between the topological qubit and  the resonator can be performed with high fidelity. Considering the coherent interaction between light and a nanoscale mechanical resonator \cite{gaoa,mejc,tkkv, kspr, evsd}, this scheme may lay the foundations for the coherent coupling between  topological and optical qubits.

\begin{figure}[t]
\includegraphics[width=8cm]{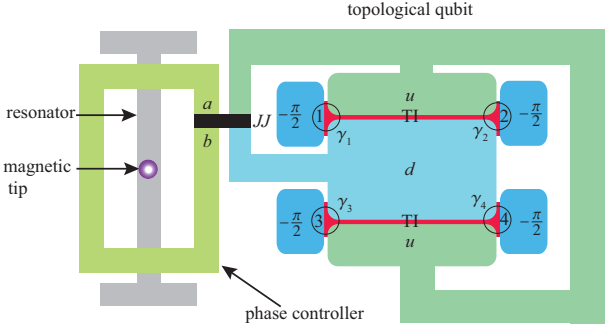}
\caption{\label{fig1}(color online). Schematics for a hybrid system comprising a topological qubit, a QPC, and a nanomechanical resonator. The topological qubit is encoded on two pairs of Majorana fermions ($(\gamma_1,\gamma_2)$ and $(\gamma_3,\gamma_4)$).  Two Majorana fermions (marked with circles) at two superconducting trijunctions are coupled though STIS quantum wire with coupling strength dependent on the phase difference between phase $\phi_u=-\pi$ of islands $u$  and phase $\phi_d=\theta$ of island $d$. The flux QPC consists of a JJ and a rf SQUID loop enclosing an external flux $\Phi_x$ which determines the phase difference  $\phi$  between superconductor islands $a$ and $b$. The resonator is covered with an array of magnetic tips. The motion of the magnetized resonator modifies the magnetic flux penetrating the plane enclosed by the QPC, resulting in  changes in the phase difference $\phi$ and in the energy splitting of the topological qubit.}
\end{figure}

{\it Hybrid system.}---The prototype hybrid  quantum system  shown in Fig.1  consists of  a topological qubit encoded on four MFs, a  QPC, and a nanomechanical resonator covered  with an array of  magnetic tips.  The flux QPC is made up of   a Josephson junction  (JJ) with two superconducting islands $a,b$ and a rf SQUID loop of inductance $L_i$ enclosing an externally applied magnetic flux $\Phi_{x}$. The phase difference $\phi$ between superconducting islands $a$ and $b$ is determined by $\phi=-2\pi \Phi_{x}/\Phi_{0}$ \cite{ljclk},   where $\Phi_{0}=h/2e$ is the flux quantum. The MFs  described by Majorana fermion operators $\gamma_i(i=1,2,3,4)$ are self-Hermitian, $\gamma_i^\dagger=\gamma_i$, and satisfy fermionic anticommutation relation $\{\gamma_i,\gamma_j\}=\delta_{ij}$.  The Majorana fermion  $\gamma_i$ is localized  at  trijunction $i(i=1,2,3,4)$, which comprises three s-wave superconductors patterned on the surface of a TI \cite{lfck}.
A pair of MFs operators $\gamma_i,\gamma_j$ can make up a Dirac fermion operator $f_{ij}=(\gamma_i-i\gamma_j)/\sqrt{2}$, which creates a fermion and $f_{ij}^\dagger f_{ij}=n_{ij}=0,1$ represents the occupation of the corresponding state. Two logical states of the topological qubit $\ket{0}_t$ and $\ket{1}_t$ are encoded on the four MFs with $\ket{0}_t=\ket{0_{12}0_{34}}$ and $\ket{1}_t=\ket{1_{12}1_{34}}$. The four MFs $\gamma_i(i=1,2,3,4)$  interacts through the superconductor- TI-superconductor (STIS) wire of width $W$, length $L$, and phases $\phi_u=-\pi$ and $\phi_d=\theta$. The effective Hamiltonian for the topological qubit reads $(\hbar=1)$ $ H_{t}=-\frac{E(\theta)}{2}\sigma^z_{t}$,
 where the coupling strength  \cite{ljck}
\be\label{eq2}
E(\theta)=\frac{v_F}{L}\sqrt{\Lambda_\theta^2+f_0^2(\Lambda_\theta)},
 \ee
 and Pauli operator $\sigma^z_{t}=(\ket{0}\bra{0}-\ket{1}\bra{1})_t$.
 Here $f_0(y)$ is the inverse function of $y=x/\tan(x)$ defined in the $0$th invertible domain, $\Lambda_\theta=\frac{\Delta_0L}{v_F}\sin\frac{\theta}{2}$ with the induced superconducting gap $\Delta_0$  and the effective Fermi velocity $v_F$ \cite{lfck}
 \be\label{eq3}
 v_F=v[\cos\mu W+\frac{\Delta_0}{\mu}\sin \mu W]\frac{\Delta_0^2}{\mu^2+\Delta_0^2},
 \ee
 where $\mu$ is the chemical potential of the TI and $v$ is the velocity of an electron on the TI's  surface.
\begin{figure}[t]
\includegraphics[width=8cm]{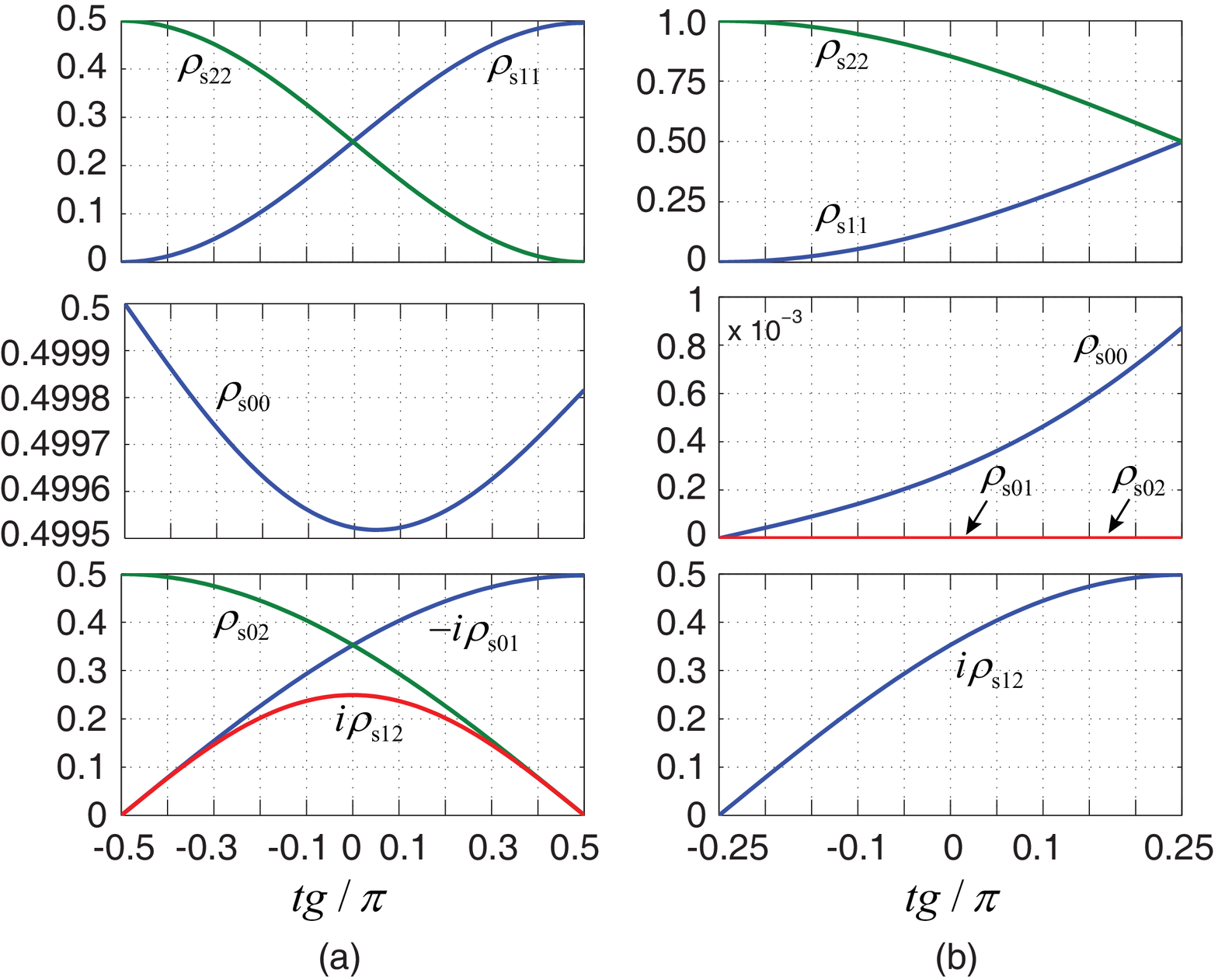}
\caption{\label{fig2}(color online). a) Numerical simulation of the state transfer:  $\frac{1}{\sqrt{2}}(\ket{\downarrow0}+\ket{\uparrow0})\rightarrow\frac{1}{\sqrt{2}} (\ket{\downarrow0}-i\ket{\downarrow1})$. The state transfer fidelity is $F_1=0.990$.  b) Numerical simulation of quantum entanglement generating, $\ket{\uparrow0}\rightarrow(\ket{\uparrow0}-i\ket{\downarrow1})/\sqrt{2}$ with a fidelity $F_2=0.993$.  The parameters used are $g=-20(2\pi)$ MHz, $g'=-100(2\pi)$ MHz, $T=25$ mK, $Q_r=1\times10^3$, $\gamma_p=1(2\pi)$ MHz, $\omega_p=4.3(2\pi)$ GHz, and $\omega_r=\omega_t=1(2\pi)$ GHz.  The corresponding matrix elements of the density matrix $\rho_s$ of the hybrid system  are $\rho_{s00}=\bra{\downarrow0}\rho_s\ket{\downarrow0}$, $\rho_{s01}=\bra{\downarrow0}\rho_s\ket{\downarrow1}$,
$\rho_{s02}=\bra{\downarrow0}\rho_s\ket{\uparrow0}$, $\rho_{s11}=\bra{\downarrow1}\rho_s\ket{\downarrow1}$,
$\rho_{s12}=\bra{\downarrow1}\rho_s\ket{\uparrow0}$, $\rho_{s22}=\bra{\uparrow0}\rho_s\ket{\uparrow0}$.}
\end{figure}

The nanomechanical resonator is described by the Hamiltonian $H_r=\omega_r a^\dagger a$ with the mechanical vibration frequency $\omega_r$ along the direction $\hat{z}$ perpendicular to the plane of area $S$ enclosed by the QPC loop, and the corresponding annihilation and creation operations $a$ and $a^\dagger$. The motion of the resonator cause a magnetic flux fluctuation $\Delta \Phi_r \simeq SGu_0(a+a^\dagger)$, where $G$ is the average magnetic field gradient produced by the magnetic tips, and $u_0$ is  the amplitude of the resonator's zero-point fluctuations. The Hamiltonian for the QPC  can be written as $H_p=\omega_p b^\dagger b$ with the plasma frequency $\omega_p\approx (CL_i)^{-1/2} $ \cite{ljclk}, and the corresponding annihilation and creation operations $b$ and $b^\dagger$.
Taking into the contribution from the magnetic tips the phase $\theta$ can be written as  \be\label{eq4}
\theta=\theta_0 +\xi\frac{a+a^\dagger}{\sqrt{2}}+\zeta\frac{b+b^\dagger}{\sqrt{2}},
\ee
 where $\theta_0$ is the corresponding phase  when the resonator is in its equilibrium position,   $\zeta\approx2\sqrt{\pi}(\frac{E_C}{E_L})^{\frac{1}{4}}$ is the magnitude of quantum fluctuations of the QPC \cite{ljclk}, and  $\xi=2\pi SGu_0/\Phi_0$.

The Hamiltonian for the whole hybrid system  described by a density matrix $\rho$ has  the form
\be\label{eq5}
H=a^\dagger a\omega_r+b^\dagger b\omega_p-\frac{1}{2}E(\theta)\sigma_t^z.
\ee
Expanding the coupling strength $E(\theta)$ to  first order in the small parameters $\frac{\xi}{\omega_r}\frac{dE(\theta)}{d\theta}|_{\theta=\theta_0}$ and $\frac{\zeta}{\omega_r}\frac{dE(\theta)}{d\theta}|_{\theta=\theta_0}$ gives the Hamiltonian
\bea \label{eq6}
H&=&a^\dagger a\omega_r+b^\dagger b\omega_p-\frac{1}{2}E(\theta_0)\sigma_t^z\notag\\
&-&\frac{1}{2}g(a^\dagger+a)\sigma_t^z- \frac{1}{2}g'(b^\dagger+b)\sigma_t^z,
\eea
where
\bea\label{eq7}
g&=&\left.\frac{\xi}{\sqrt{2}}\frac{dE(\theta)}{d\theta}\right|_{\theta=\theta_0}\notag \\  g'&=&\left.\frac{\zeta}{\sqrt{2}}\frac{dE(\theta)}{d\theta}\right|_{\theta=\theta_0}.
\eea

By rewriting Hamiltonian (\ref{eq6}) in terms of $\ket{\downarrow}=\frac{1}{\sqrt{2}}(\ket{0}+\ket{1})_t$ and  $\ket{\uparrow}=\frac{1}{\sqrt{2}}(\ket{0}-\ket{1})_t$ and applying the rotating-wave approximation and the interaction picture we obtain
\bea \label{eq9}
H_I&=&-\frac{1}{2}g(a^\dagger\sigma_t^-+a\sigma_t^+)- \frac{1}{2}g'(b\sigma_t^+e^{i(\omega_t-\omega_p)t}\notag\\
&+&b^\dagger\sigma_t^-e^{-i(\omega_t-\omega_p)t}),
\eea
 where the resonance condition $\omega_r=E(\theta_0)\equiv\omega_t$ is assumed, and   $\sigma_t^+=\ket{\uparrow}\bra{\downarrow}$  and $\sigma_t^-=\ket{\downarrow}\bra{\uparrow}$ are the raising and lowering operators, respectively.  Now we concentrate  on the experimentally relevant regime $\omega_p\gg\omega_r, g,g'$, where we can adiabatically remove the fast dynamics of the phase controller degrees of freedom.  Through projection operator techniques  we have the following Born approximation of the master equation for the reduced density matrix \cite{hbfp}:
  \be \label{eq10}
\frac{\partial}{\partial t}\rho_s(t)=-\int_{t_0}^tdt'\text{Tr}_p[H_I(t),[H_I(t'),\rho_s(t')\otimes\rho_p]],
\ee
  where $\rho_p$ is the steady state of the QPC in the absence of the qubit-resonator system. We perform  the Markov approximation on equation (\ref{eq10}) by replacing $\rho_s(t')$ with $\rho_s(t)$ and by sending $t_0\rightarrow -\infty$, resulting in the Markovian quantum master equation
  \be \label{eq10a}
\frac{\partial}{\partial t}\rho_s(t)=-\int_{0}^\infty d\tau\text{Tr}_p[H_I(t),[H_I(t-\tau),\rho_s(t)\otimes\rho_p]],
\ee
 This Markov approximation holds if the QPC modes decay much faster than $g'^{-1}$ or if they are far detuned from the topological qubit by much more than $g'^{-1}$ \cite{kspr}.
   Substituting $H_I$(\ref{eq6}) into equation \eqref{eq10} gives (neglecting transients by dispatching $t_0\rightarrow -\infty$)
 \bea \label{eq11}
\frac{\partial}{\partial t}\rho_s&=&-\frac{1}{4}g'^2\left[J(\omega_t)(\sigma_t^+\sigma_t^-\rho_s-\sigma_t^-\rho_s\sigma_t^ +)\right.\notag\\ &+&\left.K(\omega_t)(\sigma_t^-\sigma_t^+\rho_s-\sigma_t^+\rho_s\sigma_t^-)+\text{H.c.}\right],
\eea
where
\be \label{eq12}
J(\omega_t)=\int_0^\infty \avg{b(\tau)b^\dagger(0)}e^{i\omega_t\tau}d\tau
\ee

\be \label{eq13}
K(\omega_t)=\int_0^\infty \avg{b^\dagger(\tau)b(0)}e^{-i\omega_t\tau}d\tau
\ee

To describe dissipative effects we introduce the quantum Langenvin equation for the QPC degrees of freedom in the limit $g'\rightarrow 0$:
\be\label{eq8}
\dot{b}=-i[b, H_p]-\frac{\gamma_p}{2} b-\sqrt{\gamma_p}\varsigma
\ee
where the noise operator $\varsigma$ fulfills $\avg{ \varsigma^\dagger(t)\varsigma(t')}=N_p\delta(t-t')$  with  $N_p=[\text{exp}(\omega_p/k_BT)-1]^{-1}$ and $\gamma_pN_p$ is the relevant decoherence rate.
From QLE (\ref{eq8}) by Fourier transformation the steady-state correlation functions $J(\omega_t)$ and $K(\omega_t)$ can be obtained as
\be \label{eq14}
J(\omega_t)=\frac{\gamma_p(N_p+1)}{2\omega_p}[(i\omega_t+i\omega_p-\frac{\gamma_p}{2})^{-1} -(i\omega_t-i\omega_p-\frac{\gamma_p}{2})^{-1}],
\ee
\be \label{eq15}
K(\omega_t)=\frac{\gamma_p N_p}{2\omega_p}[(-i\omega_t+i\omega_p-\frac{\gamma_p}{2})^{-1} -(-i\omega_t-i\omega_p-\frac{\gamma_p}{2})^{-1}]
\ee
Rewriting  equation \eqref{eq11} gives the following effective master equation
 \bea \label{eq16}
 \frac{\partial}{\partial t}\rho_s&=&-i\frac{\Delta}{2}[\sigma_t^z,\rho_s]+\Gamma_p(N_p+1) D(\sigma_t^-)\rho_s +\Gamma_p N_p D(\sigma_t^+)\rho_s \notag\\
  &+&\gamma_r(N_r+1) D(a)\rho_s+\gamma_r N_r D(a^\dagger)\rho_s,
\eea
 where we have included the dissipation of the resonator modes for a mechanical quality factor $Q_r=\omega_r/\gamma_r$, $D[\hat{c}]\rho_s:=(2\hat{c}\rho_s\hat{c}^\dagger-\hat{c}^\dagger \hat{c}\rho_s-\rho_s \hat{c}^\dagger \hat{c})/2$, $N_r=[\text{exp}(\omega_r/k_BT)-1]^{-1}$, $\Delta=\frac{\gamma_pg'^2}{2\omega_p^2}$, and $\Gamma_p=\frac{2\gamma_p^2\omega_tg'^2}{\omega_p^4}$.

 {\it Example.}---As an example we discuss a SiC  beam  of dimensions $(l,w,t)=(1.1,0.12,0.075) \mu$m with a  basic mode of frequency $1(2\pi)$ GHz, $u_0\approx 15$ fm, and  $Q=500$ at temperature $T=4.2$ K \cite{mphj, xmhh}, or $Q\approx2300$ at $T=25$ mK according to the temperature dependence of the quality factor $Q^{-1}\propto T^{0.3}$ \cite{xfcz}. A magnetic tip of size of 50 nm  with homogeneous magnetization $M\approx2.3\times10^{6}$ \cite{hmmp,prpc} attached on the resonator produces a magnetic gradient of $ G \approx1\times10^{8} $  T$/$m at a distance of 1 $\mu$m , resulting in  $\xi>0.002$ for a surface $S\approx1\mu$m$^2$. The QPC comprises a large Josephson junction  \cite{jmsn} and a rf SQUID loop with very small inductance \cite{jfvp},  we may set $\zeta\approx 0.01$ and $\omega_p\approx 4.3(2\pi)$ GHz \cite{ljclk}. For topological qubit we may choose $\Delta_0\approx 25(2\pi)$ GHz \cite{vmou}, $L\sim5 \mu$m, and $v_F\approx2.2\times10^4$ m/s by adjusting the TI's chemical potential $\mu$ (\ref{eq3}). From equations (\ref{eq2}, \ref{eq7}) we obtain $g\approx-20(2\pi)$ MHz and $g'\approx-100(2\pi)$ MHz for  $\theta_{\text{on}}=0.09$;  $g\approx -5$ KHz and $g'\approx-25$ KHz   for  $\theta_{\text{off}}=3.1$

 {\it Applications.}---The coupling strength $g$ can be coherently controlled by modifying the phase $\theta$: the interaction between the qubit and the resonator is switched on ( off) by tuning $\theta$ to $\theta_{\text{on}}$ ( $\theta_{\text{off}}$ ). A unitary transformation
\be
\mu\ket{\downarrow0}+\nu\ket{\uparrow0}\rightarrow \mu\ket{\downarrow0}-i\nu\ket{\downarrow1},
\ee
can be performed by adiabatically turn on the coupling for a duration corresponding to a $\pi$ pulse $\int g(t)dt=-\pi$. Next a single-qubit rotation on the latter can then finish a quantum state transfer from the topological qubit to the motion mode of the resonator, where $\mu$ and $\nu$ are arbitrary complex numbers satisfying $|\mu|^2+|\nu|^2=1$. A maximally entangled state $\ket{\uparrow0}\rightarrow(\ket{\uparrow0}-i\ket{\downarrow1})/\sqrt{2}$ can be generated if $\int g(t)dt=-\pi/2$. The choose of  $\int g(t)dt=-3\pi/2$ accomplishes a $\sqrt{\text{SWAP}}$ gate, the squared root of SWAP gate, up to a single-qubit rotation. Series of  $\sqrt{\text{SWAP}}$ gates and single-qubit $90^\circ$ rotations  about $\hat{z}$  on the subsystem $i$  denoted by $\text{R}_{z,i}(90)$ gives the controlled-phase ($\text{CP}_{t,r}$) gate
\be
\text{CP}_{t,r}=\text{R}_{z,t}(90)\text{R}_{z,r}(-90)\sqrt{\text{SWAP}} \text{R}_{z,t}(180) \sqrt{\text{SWAP}}
 \ee
 for the hybrid system. Finally  an arbitrary unitary transformation on the hybrid system can be decomposed into  $\text{CP}_{t,r}$ gates and single-qubit rotations \cite{mnic}.
\begin{figure}[t]
\includegraphics[width=8cm]{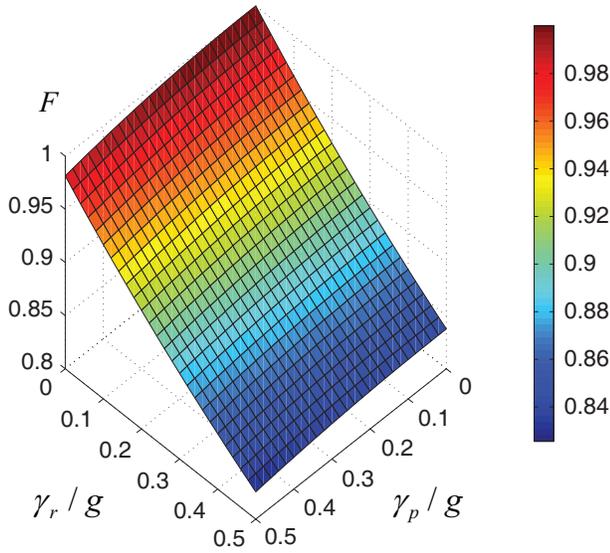}
\caption{\label{fig3}(color online). The effect of decoherence sources $\gamma_r$ and $\gamma_p$ on the fidelity of state  transfer: $\frac{1}{\sqrt{2}}(\ket{\downarrow0}+\ket{\uparrow0})\rightarrow\frac{1}{\sqrt{2}} (\ket{\downarrow0}-i\ket{\downarrow1})$. Other parameters are as in Fig.\ref{fig2}.  }
\end{figure}

{\it Numerical simulations.}---The main sources of error of the quantum manipulations discussed above are decoherence from the resonator and the QPC.  Low temperature is required to exponentially decrease the probability of the occupation of the excitation modes of the STIS wire by the factor $\gamma_w\equiv\text{exp}(\frac{-v_F}{k_BTL})$ \cite{ljck}: $T=20$ mK gives $\gamma_w<10^{-3}$ for the aforesaid values of $v_F$ and $L$. The finite length of STIS wire will have very small effect on  the coherence time of the topological qubit itself: the energy splitting $E(\theta_{\text{off}})\sim1(2\pi)\times10^3$ Hz for $\phi_{\text{off}}=0.13$. The energy splitting $E(\theta_{\text{on}})$ may be affected by some processes, such as dynamics modulations of the superconducting gap and variation of the electromagnetic environment.

The error of the quantum information transfer between the topological qubit and the resonator is estimated in terms of fidelity by numerical solving  the effective  master equation \eqref{eq16}. We may choose $\omega_r=E(\theta_{\text{on}})=\omega_t=1(2\pi)$ GHz, $\omega_p=4.3(2\pi)$ GHz, $T=20$ mK \cite{acmh},  $\gamma_p=1$ MHz \cite{jcfw}, $Q_r=2\times10^3$,   $g=-20(2\pi)$ MHz, and  $g'/2\pi=100(2\pi)$ MHz.
The evolution of the state transfer
\be\label{eq}
\frac{1}{\sqrt{2}}(\ket{\downarrow0}+\ket{\uparrow0})\xrightarrow{\int^{t_{f1}} g(t)dt=-\pi}\ket{\psi_1}\equiv\frac{1}{\sqrt{2}}(\ket{\downarrow0}-i\ket{\downarrow1})
 \ee
 and the generating of a maximally entangled state
 \be\label{eq}
 \ket{\uparrow0}\xrightarrow{\int^{t_{f2}} g(t)dt=-\pi/2}\ket{\psi_2}\equiv(\ket{\uparrow0}-i\ket{\downarrow1})/\sqrt{2}
  \ee
  are shown in Fig. \ref{fig2}a) and b), respectively, with the corresponding fidelity $F_1=\bra{\psi_1}\rho_s(t_{f1})\ket{\psi_1}=0.990$ and $F_2=\bra{\psi_2}\rho_s(t_{f2})\ket{\psi_2}=0.993$. The influence of the decoherence sources $\gamma_r$ and $\gamma_p$ on the state transfer fidelity $F_1$ is shown in  Fig.\ref{fig3}.
Finally we estimate the influence of the fluctuations in the energy splitting $E(\theta_{\text{on}})$ on the operation fidelity by assuming unknown errors in $E(\theta_{\text{on}})$, and $g$: the corresponding fidelity $F_1$ decreases from 0.989 to 0.984 for 1\% unknown errors in $E(\theta_{\text{on}})$ and $g$.

{\it Conclusion.}---In summary, we have presented a scheme for quantum information transfer between topological qubit and the quantized motion of a nanomechanical resonator. Quantum state transfer, quantum entanglement generating, and arbitrary unitary transformation in the topological-qubit-resonator system may be performed with high fidelity. Considering the advances in coherent transfer of quantum information between the quantized motion of the resonator and other conventional qubits including optical qubits \cite{acmh, kspr, prpc,mljs,tkkv}, this quantum interface enables us to store conventional quantum information on topological qubits for long time storage, to efficiently detect  topological qubit states, to design partially protected universal topological quantum computation, where topological qubit can receive a single-qubit state prepared by a conventional qubit with high accuracy,  compensating the topological qubit's incapability of generating some single-qubit states.

 This work was supported by the National Natural Science Foundation of China ( 11072218 and 11272287), by Zhejiang Provincial Natural Science Foundation of China (Grant No. Y6110314), and  by Scientific Research Fund of Zhejiang Provincial Education Department (Grant No. Y200909693).

\end{document}